\documentclass[aps,prc,amsfonts,preprintnumbers,superscriptaddress,showpacs,nofootinbib]{revtex4}

\usepackage{graphics}
\usepackage{psfrag}
\usepackage{epsfig}
\usepackage{epsf}
\usepackage{float}

\newcommand{\fet}[1]{\mbox{\boldmath $#1$}}

\newcommand{\beq}{\begin{equation}}
\newcommand{\eeq}{\end{equation}}
\newcommand{\beqa}{\begin{eqnarray}}
\newcommand{\eeqa}{\end{eqnarray}}
\newcommand{\nn}{\nonumber \\ }

\begin{document}

\title{Subleading contributions to the chiral three-nucleon 
force II:\\[0.2em]
short-range terms and relativistic corrections}

\author{V.~Bernard}
\email[]{Email: bernard@ipno.in2p3.fr}
\affiliation{Institut de Physique Nucl\'eaire, CNRS/Univ. Paris-Sud
  11, (UMR 8608), F-91406 Orsay Cedex, France}
\author{E.~Epelbaum}
\email[]{Email: evgeny.epelbaum@rub.de}
\author{H.~Krebs}
\email[]{Email: hkrebs@itkp.uni-bonn.de}
\affiliation{Institut f\"ur Theoretische Physik II, Ruhr-Universit\"at Bochum,
  D-44780 Bochum, Germany}
\author{Ulf-G.~Mei{\ss}ner}
\affiliation{Universit\"at Bonn, Helmholtz-Institut f{\"u}r
  Strahlen- und Kernphysik (Theorie) and Bethe Center for Theoretical Physics, 
  D-53115 Bonn, Germany}
\affiliation{Forschungszentrum J\"ulich, Institut f\"ur Kernphysik 
  (IKP-3), Insitute for Advanced Simulation (IAS-4) and J\"ulich Center
   for Hadron Physics, D-52425 J\"ulich, Germany}
\email[]{Email: meissner@itkp.uni-bonn.de}
\homepage[]{URL: www.itkp.uni-bonn.de/~meissner/}
\date{\today}

\begin{abstract}
We derive the short-range contributions and the leading relativistic
corrections to the three-nucleon force at
next-to-next-to-next-to-leading order in the chiral expansion.
\end{abstract}


\maketitle

\vspace{-0.2cm}

\section{Introduction}
\def\theequation{\arabic{section}.\arabic{equation}}
\label{sec:intro}

Chiral effective field theory allows for a consistent calculation of 
forces between  two, three and more nucleons (for recent reviews, 
see Refs.~\cite{Epelbaum:2008ga,Machleidt:2011zz,KalantarNayestanaki:2011wz}).
In this paper, we fill in the last missing part of the nuclear forces 
calculated at next-to-next-to-next-to-leading order (N$^3$LO) in the
chiral expansion, namely the
short-range contributions and the leading relativistic corrections 
to the three-nucleon force (3NF). At this order,
the 3NF is given by five topologies as shown in Ref.~\cite{Bernard:2007sp}.
Three of these do not involve multi-nucleon operators and thus
contribute at long range $r \sim 1/M_\pi$ and $r \sim 1/(2M_\pi)$, with $M_\pi$ the pion mass.  
The corresponding momentum and
coordinate-space representations are given in that paper (see also
\cite{Ishikawa:2007zz}). Here, we
work out the remaining terms corresponding to the two topologies involving
four-nucleon operators (the  short-range terms) as well as the
relativistic ($1/m$) corrections to the leading (N$^2$LO) one- and two-pion 
exchange contributions. Throughout, $m$ denotes the nucleon mass. 
Special care has to be taken to calculate these
corrections consistently with the ones contributing to the
two-nucleon force. With the formalism given here, one is now in the position
to perform calculations in three (or more) nucleon systems consistently
at N$^3$LO. It is important to stress that the 3NF at this order does {\sl not}
involve any new and unknown low-energy constants (LECs) - the full 3NF to this 
order thus depends just on two LECs, commonly called $D$ and $E$, that
parameterize the leading  one-pion-contact term topology and the six-nucleon
contact term at N$^2$LO. Note that an exploratory study of the effects of the long-range
parts of the 3NF at N$^3$LO in the triton was recently presented in~\cite{Skibinski:2011vi}.

Our manuscript is organized as follows: In Sec.~\ref{sec2} we discuss  
the contributions of the one-pion-exchange-contact topology. Sec.~\ref{sec3}
describes the calculation of the two-pion-exchange-contact topology
while Sec.~\ref{sec4} is devoted to the analysis of the leading
relativistic corrections  that  also
appear at N$^3$LO. 
We end with a brief summary and outlook. For the sake of completeness,
in appendix \ref{app1} we give the formal algebraic structure of the
parts of the nuclear Hamiltonian which are relevant for our
work. Finally, appendix \ref{app2} contains the coordinate-space
representation of the obtained results.

\section{The one-pion-exchange-contact topology}
\def\theequation{\arabic{section}.\arabic{equation}}
\label{sec2}

\begin{figure}[tb]
\vskip 1 true cm
\includegraphics[width=15.0cm,keepaspectratio,angle=0,clip]{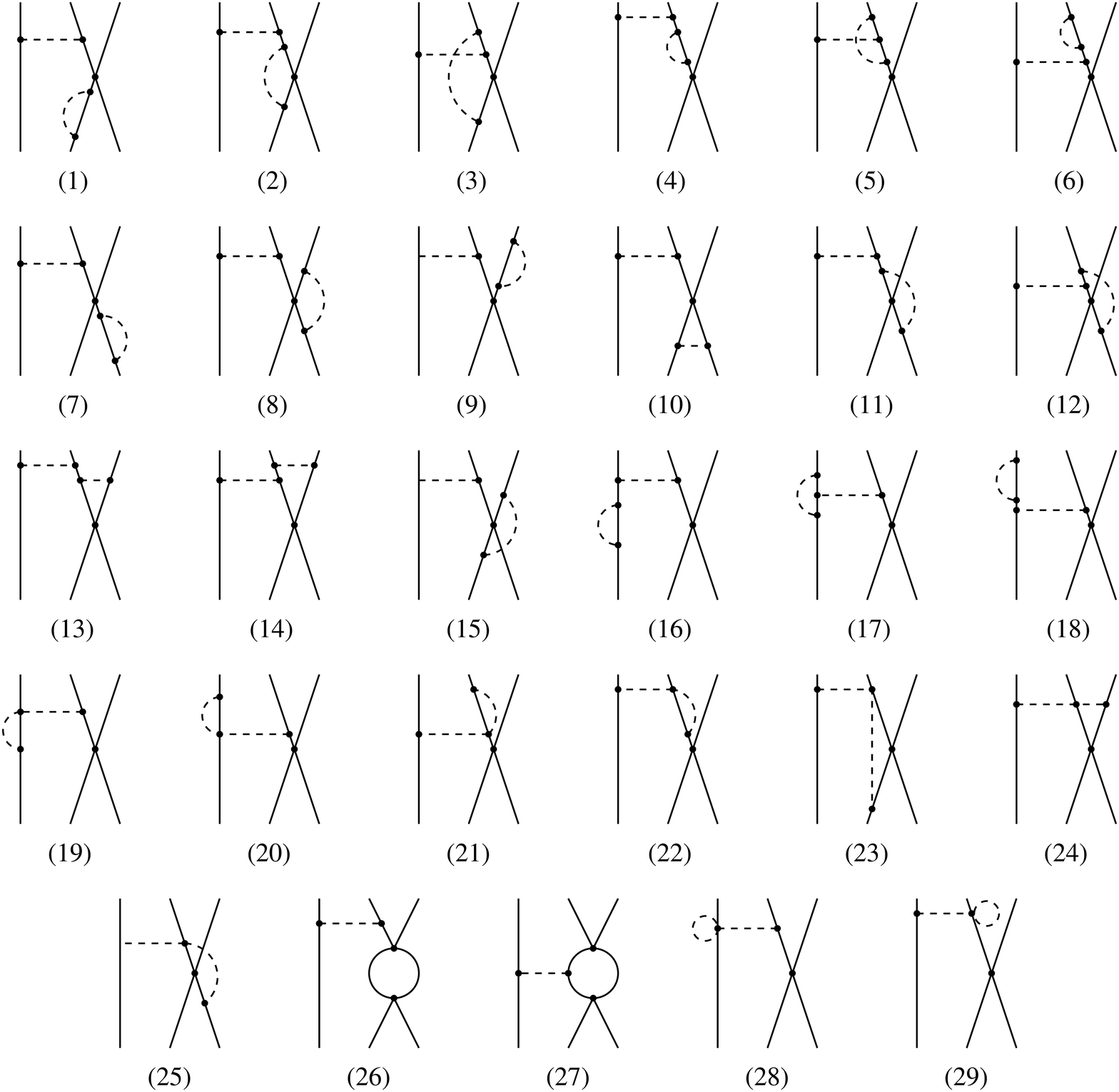}
    \caption{
         Diagrams contributing to the one-pion-exchange-contact
         3NF. Solid and dashed lines
         represent nucleons and pions, respectively. Solid dots refer to the
         lowest-order  (dimension one) vertices
         from the chiral Lagrangian.  Diagrams which result from the
         interchange of the nucleon lines and/or
         application of the time reversal operation are not shown. 
\label{fig1} 
 }
\end{figure}

We begin with the discussion of the one-pion-exchange contributions
involving short-range contact interactions between two nucleons which
are shown  in Fig.~\ref{fig1}. It is important to keep in mind that
these diagrams do not correspond to Feynman graphs. Rather,
they represent schematically various contributions to the connected,
irreducible part of the three-nucleon amplitude which gives rise to
the three-nucleon force. 
Each diagram is to be understood as a set of all possible
time-ordered-like graphs of the same topology (i.e.~the same sequence of
vertices). The precise meaning of the diagrams and their
contributions to the nuclear Hamiltonian depend, strictly speaking, 
on the approach used to extract the irreducible part of the
amplitude.\footnote{We remind the reader that nuclear potentials
  do not correspond to observable quantities and are, in general,  scheme
  dependent. } 
In this work, we
adopt the method of unitary transformation which was already successfully 
applied to the derivation of nuclear forces  
Ref.~\cite{Epelbaum:1998ka,Epelbaum:1999dj,Epelbaum:2002gb,Epelbaum:2004fk,Epelbaum:2007us} and 
more recently electromagnetic nuclear exchange current and charge
operators \cite{Kolling:2009iq,Kolling:2011mt}. It is also the same method that
we used in Ref.~\cite{Bernard:2007sp} to compute the long-range parts
of the 3NF at N$^3$LO. The resulting
two- and many-body forces as well as the exchange currents are
thus derived in the same theoretical approach from the effective
chiral Lagrangian and are, per construction, \emph{consistent} with
each other. 

Before discussing the results for individual diagrams we first
outline the way the actual calculation is carried out. Starting from
the effective chiral Lagrangian for pions and nucleons, we first
compute the corresponding Hamilton density using the canonical
formalism, see \cite{Gerstein:1971fm} for more details. In order to
arrive at nuclear forces (and current operators) valid in the
low-energy region, i.e.~well below the pion production threshold, the
pion fields need to be integrated out. This is achieved through 
decoupling of the purely nucleonic subspace of the Fock space from the
rest via a suitably chosen unitary transformation (UT)
\cite{Epelbaum:1998ka,Epelbaum:1999dj,Epelbaum:2002gb}. This 
step is carried out perturbatively utilizing the chiral expansion and
making use of the standard chiral power counting based on naive
dimensional analysis. We also exploit the freedom to choose the basis
states in the nucleonic subspace of the Fock space and include a large
number of additional unitary transformations which can be constructed
at the given chiral order \cite{Epelbaum:2007us}. It turns out that the resulting unitary
ambiguity of the nuclear Hamiltonian and exchange currents is
strongly constrained by requiring that the resulting nuclear
potentials and currents are renormalizable, see \cite{Epelbaum:2007us}
for a detailed discussion. The explicit form of the (purely strong)
unitary transformation compatible with the renormalizability
requirement and the operators contributing to effective nuclear
Hamiltonian which are needed in the present calculation can be found in
appendix \ref{app1}, see also
Ref.~\cite{Epelbaum:2007us}. Once the unitary operator is determined at the
desired order in the chiral expansion, the effective, purely nucleonic
Hamiltonian can be computed straightforwardly. One ends up with
terms given by a sequence of vertices and energy denominators which
are similar to the ones emerging in time-ordered perturbation
theory. However, the energy denominators and the coefficients in front of each
terms generally differ from the ones arising in the
context of time-ordered perturbation theory. We refer to \cite{Epelbaum:2005pn} for
more details on the method of unitary transformation. 

We now discuss the individual contributions to the three-nucleon
force. Unless not stated otherwise, the expressions for the 3N potentials
are to be understood as matrix elements with respect to the nucleon
momenta while as operators with respect to spin and isospin quantum
numbers. We begin with diagrams (1)-(6) in Fig.~\ref{fig1} with one  
self-energy/vertex-correction insertion on the  second nucleon. 
The contribution of these diagrams to the 3NF depends on the 
unitary transformations given in Eqs.~(3.24), (3.25) and (3.48) of
Ref.~\cite{Epelbaum:2007us}. Fixing the corresponding rotation
angles $\alpha_{1,2,6}$ in the way compatible with renormalizability
as described in that work leads to the following result for the
corresponding 3NF contribution
\beqa
\label{cont1}
V &=& \frac{g_A^4 C_T}{ 4 F_\pi^4} \fet \tau_1 \cdot \fet \tau_2
\frac{\vec \sigma_1 \cdot \vec q_1}{q_1^2 + M_\pi^2} 
\int \frac{d^3l}{(2 \pi)^3} \, \frac{1}{\omega_l^4} \left[
- (\vec q_1 \cdot \vec l )( \vec l \cdot \vec \sigma_3 )+ (\vec q_1 \cdot
\vec \sigma_3 ) \vec l^2 \right] \nn
&\stackrel{\rm DR}{=} & - \frac{g_A^4 C_T}{ 16 \pi  F_\pi^4} M_\pi \, \fet \tau_1 \cdot \fet \tau_2
\, \frac{(\vec \sigma_1 \cdot \vec q_1)(\vec \sigma_3 \cdot \vec q_1)}{q_1^2 + M_\pi^2} \,,
\eeqa
where $\vec \sigma_i$ ($\fet \tau_i$) refer to the Pauli spin
(isospin) matrices while $\vec q_i \equiv \vec p_i \, ' - \vec p_i$
denote the momentum transfer of the nucleon $i$. Further, $g_A$ and
$F_\pi$ stay for the nucleon axial-vector and pion decay coupling
constant, respectively, while $C_T$ is the low-energy constant
accompanying the lowest-order (spin-dependent) two-nucleon contact
interaction. We emphasize that the Wigner-symmetry-invariant 
two-nucleon contact interaction proportional to $C_S$ yields a vanishing  
contribution which can be traced back to the fact  that it
commutes with the pion-nucleon vertex.  Here and in what follows, we
use dimensional regularization (DR) in the derivation of the
3NF. Further, we adopt the same notation as in Ref.~\cite{Bernard:2007sp} and give results for a
particular choice of nucleon labels (unless not stated otherwise). The
full expression for the 3NF results by taking into account all
possible permutations of the nucleons\footnote{For three nucleons
  there are altogether 6 permutations.}, i.e.:
\beq
\label{convention}
V_{\rm 3N}^{\rm full} = V_{\rm 3N} \; +\;\mbox{all permutations}\, .
\eeq 
Next, the contribution of diagrams (7)-(9) in Fig.~\ref{fig1}  with one  
self-energy/vertex-correction insertion on the third nucleon is found
to vanish completely. This feature does not depend on the structure of
vertices appearing in these diagrams and emerges from the
renormalizability constraints, see Ref.~\cite{Epelbaum:2007us}.  
On the other hand, diagrams (10)-(15) with pions being exchanged
between the nucleons 1 and 2 and 2 and 3 do yield a nonvanishing
contribution of the form
\beqa
\label{cont2}
V &=& \frac{g_A^4 C_T}{ 4 F_\pi^4} 
\frac{\vec \sigma_1 \cdot \vec q_1}{q_1^2 + M_\pi^2} 
\int \frac{d^3l}{(2 \pi)^3} \, \frac{1}{\omega_l^4} \left\{
\fet \tau_1 \cdot \fet \tau_3 \left[ (\vec q_1 \cdot \vec l )( \vec l
  \cdot \vec \sigma_3 ) - (\vec q_1 \cdot \vec \sigma_3 \, ) l^2
\right] + [\fet \tau_1 \times \fet \tau_2 ] \cdot \fet \tau_3 \, \vec
l \cdot [ \vec \sigma_2 \times \vec \sigma_3 ] \, \vec q_1 \cdot \vec l
\, \right\}
\nn
&\stackrel{\rm DR}{=} & \frac{g_A^4 C_T}{ 32 \pi  F_\pi^4} M_\pi 
\, \frac{\vec \sigma_1 \cdot \vec q_1}{q_1^2 + M_\pi^2} 
\Big[
2 \fet \tau_1 \cdot \fet \tau_3 (\vec q_1 \cdot \vec \sigma_3 ) 
 - \fet \tau_1 \cdot [\fet \tau_2  \times \fet \tau_3 ]\, \vec
q_1 \cdot [ \vec \sigma_2 \times \vec \sigma_3 ]  \Big]
\,.
\eeqa
We find that diagrams (16)-(18) involving a single
self-energy/vertex-correction insertion on the nucleon 1 do not
generate 3NF contributions. For graph (17) this feature is quite
general and does not depend on the vertex structure. Irregardless of
the choice for the additional unitary transformations, this diagram
appears to be purely reducible at the order considered. Its
contribution to the scattering amplitude is, therefore, properly
accounted for by iterating the dynamical equation. The two self-energy
corrections do yield non-vanishing irreducible contributions of
opposite sign  when considered separately. Since the
leading nucleon self-energy correction is scalar, isoscalar and independent of
the nucleon momentum, the contributions of these two diagrams sum up
to zero. 

Similarly, diagrams (19)-(25) proportional to $g_A^2 C_{S,T}$ do not
produce any non-vanishing 3NF contributions. This is because the
integrands entering the corresponding loop integrals are always 
odd functions of the loop momentum. This feature depends crucially
on the renormalizability constraints discussed above which ensure that
the pion-exchange between the  nucleons 1 and 2 factorizes out in all
irreducible contributions emerging from these diagrams. Thus, the pion
loop integrals are always proportional to $\omega_l^{-2}$ (which can
be understood from the dimensional analysis) and involve a single
power of the loop momentum $\vec l$ in the numerators emerging from
the leading (derivative) pion-nucleon vertex $\sim g_A$.      

\begin{figure}[tb]
\vskip 1 true cm
\includegraphics[width=7.0cm,keepaspectratio,angle=0,clip]{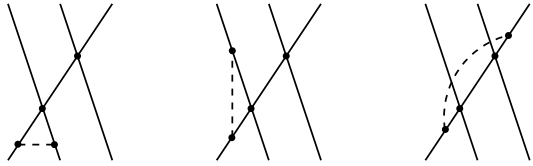}
    \caption{
         Examples of diagrams $\propto g_A^2 C_{S,T}^2$ which lead to finite
         shifts of the purely short-range 3NF at N$^2$LO as explained
         in the text. For notation see Fig.~\ref{fig1}.  
\label{fig4} 
 }
\end{figure}

Next,  diagrams (26), (27) in Fig.~\ref{fig1} proportional to
$g_A^2 C_{S,T}^2$  also lead to
vanishing 3NF contributions for the choices of the additional UTs
compatible with the renormalizability constraints.\footnote{To avoid 
possible confusion, we emphasize once again that these diagrams
correspond to all possible non-iterative time-ordered graphs wich
nucleons treated as static sources. Non-static corrections are
computed perturbatively as discussed in section \ref{sec4}.} 
It should be emphasized that there are also purely short-range
contributions to the 3NF $\propto g_A^2 C_{S,T}^2$ emerging from
diagrams shown in Fig.~\ref{fig4} with NN contact interactions acting between
\emph{different} pairs of nucleons.   
While these graphs (presumably) generate non-vanishing 3NFs, their
contributions are purely short-range and only provide a finite
shift $\propto M_\pi$ 
to the LEC $E$ which accompanies the contact 3NF at N$^2$LO.  Since we
anyway work with bare LEC $E$ which needs to be refitted to
experimental data at each order in the chiral expansion, there is no
need to explicitly evaluate these contributions unless one is  interested in the
quark mass dependence of the 3NF and few-nucleon  observables.  

\begin{figure}[tb]
\vskip 1 true cm
\includegraphics[width=15.0cm,keepaspectratio,angle=0,clip]{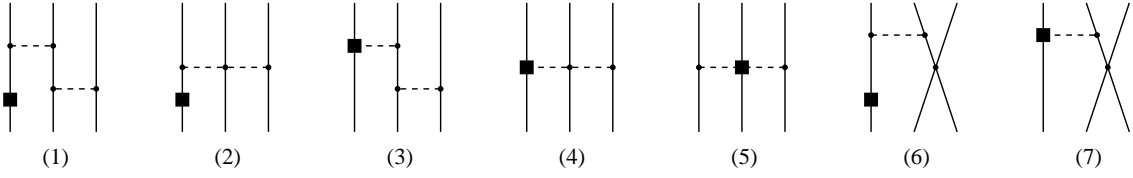}
    \caption{
         Tree diagrams contributing to the two-pion-exchange and
         one-pion-exchange-contact topology of the 3NF at N$^3$LO. 
         The solid boxes denote insertions of either sub-subleading
         $d_i$-vertices from the effective pion-nucleon 
         Lagrangian or the leading $1/m$-corrections. For notation see Fig.~\ref{fig1}.  
\label{fig3} 
 }
\end{figure}

Finally, the last two
diagrams in Fig.~\ref{fig1}  clearly do not yield any 
contributions in a complete analogy with NLO one-pion-exchange-contact
diagrams. Finally, there are also no contributions at N$^3$LO from tree diagrams
involving one insertion of the higher-order $d_i$-vertices in the effective
Lagrangian, see graphs (6) and (7) in Fig.~\ref{fig3} except for the
relativistic corrections which will be considered in section
\ref{sec4}. As explained in Ref.~\cite{Bernard:2007sp}, the contributions from these
diagrams are suppressed by at least one power of $Q/m$ where $Q$ denotes
                 a genuine soft scale.  

We are thus left
with Eqs.~(\ref{cont1}), (\ref{cont2}) as the only nonvanishing
contributions to the one-pion-exchange-contact 3NF topology. We now
show that these terms cancel each other exactly if one takes into account the
antisymmetric nature of few-nucleon states. In particular, we use the
identities 
\beqa
\Big( \fet \tau_3 \sigma_3^i + \fet \tau_2 \sigma_2^i  \Big)_{\rm A23}
&=& \frac{1}{4} \left( \fet \tau_3 \sigma_3^i + \fet \tau_2 \sigma_2^i  
- \fet \tau_3 \sigma_2^i - \fet \tau_2 \sigma_3^i  + \fet \tau_2
\times \fet \tau_3 \, [ \vec \sigma_2 \times \vec \sigma_3 \,]^i
\right) \nn
&\equiv & \fet B^i \,, \nn
\Big( \fet \tau_2
\times \fet \tau_3 \, [ \vec \sigma_2 \times \vec \sigma_3 \,]^i  \Big)_{\rm A23}
&=& 2  \fet B^i \,, \nn
\Big( \fet \tau_3 \sigma_2^i + \fet \tau_2 \sigma_3^i  \Big)_{\rm A23}
&=& -  \fet B^i \,,
\eeqa
where the superscript $i$ refers to the cartesian component of the Pauli
spin matrices and $\rm A23$ denotes antisymmetrization with respect to nucleons
2 and 3, which,  for a momentum-independent operator $X$,  can be written in the form   
\beq 
( X )_{\rm A23} \equiv \frac{1}{2} \left( X - \frac{1 + \vec
    \sigma_2 \cdot \vec \sigma_3}{2}\, \frac{1 + \fet \tau_2 \cdot
    \fet \tau_3}{2} X \right)\,.
\eeq
It is easy to see that adding the contribution from interchanging
the nucleons 2 and 3 to Eqs.~(\ref{cont1}) and (\ref{cont2}) and
performing antisymmetrization with respect to these nucleons leads to
a vanishing result. We, therefore, conclude that there are {\sl no}
one-pion-exchange-contact terms in the 3NF at N$^3$LO.

\section{The two-pion-exchange-contact topology}
\def\theequation{\arabic{section}.\arabic{equation}}
\label{sec3}

\begin{figure}[tb]
\vskip 1 true cm
\includegraphics[width=15.0cm,keepaspectratio,angle=0,clip]{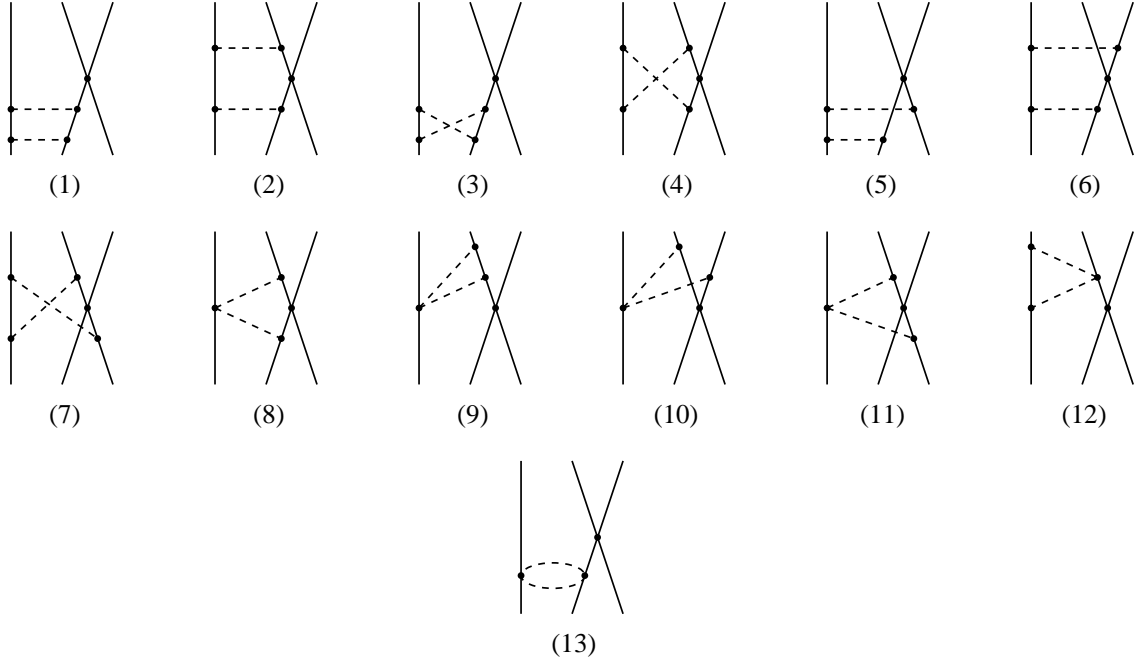}
    \caption{
         Diagrams contributing to the two-pion-exchange-contact topology of the 3NF. 
         For notation see Fig.~\ref{fig1}.  
\label{fig2} 
 }
\end{figure}

We now turn to the two-pion-exchange-contact diagrams shown in
Fig.~\ref{fig2}. Evaluating the  matrix elements of the operators
listed in Eq.~(\ref{class4}) 
for diagrams (1)-(7)
in this figure we find the $g_A^4 C_{S,T}$-contribution to the
two-pion-exchange-contact topology of the form
\beqa
\label{2picont1}
V_{\rm 2\pi-cont} &=&  \frac{g_A^4 C_T}{ 8 F_\pi^4} 
\int \frac{d^3l}{(2 \pi)^3} \,\bigg\{ 
\fet \tau_1 \cdot \fet \tau_2 \bigg[ \bigg( \frac{1}{\omega_+^4
  \omega_-^2}  + \frac{1}{\omega_+^2 \omega_-^4} \bigg) \Big( q_1^2 (\vec q_1 \cdot \vec
\sigma_2 \, ) (\vec q_1 \cdot \vec \sigma_3 \, )  +  2 q_1^2 l^2 (\vec
\sigma_2  \cdot \vec \sigma_3 \, ) 
-q_1^2 (\vec l \cdot \vec
\sigma_2 \, ) (\vec l \cdot \vec \sigma_3 \, )  \nn
&& {} - q_1^4 ( \vec
\sigma_2  \cdot \vec \sigma_3 \, )  - l^2 (\vec q_1 \cdot \vec
\sigma_2 \, ) (\vec q_1 \cdot \vec \sigma_3 \, ) + 
l^2 (\vec l \cdot \vec
\sigma_2 \, ) (\vec l \cdot \vec \sigma_3 \, ) - l^4 ( \vec
\sigma_2  \cdot \vec \sigma_3 \, )    
\Big) + \bigg( \frac{1}{\omega_+^4
  \omega_-^2}  - \frac{1}{\omega_+^2 \omega_-^4} \bigg) \nn
&& {} \times\Big( - q_1^2 (\vec q_1 \cdot \vec
\sigma_2 \, ) (\vec l \cdot \vec \sigma_3 \, ) + q_1^2 (\vec l \cdot \vec
\sigma_2 \, ) (\vec q_1 \cdot \vec \sigma_3 \, ) 
 + l^2 (\vec q_1 \cdot \vec
\sigma_2 \, ) (\vec l \cdot \vec \sigma_3 \, ) - l^2 (\vec l \cdot \vec
\sigma_2 \, ) (\vec q_1 \cdot \vec \sigma_3 \, ) 
\Big)
\bigg]  \nn
&&{} + 2 \fet \tau_2 \cdot \fet \tau_3    \bigg( \frac{1}{\omega_+^4
  \omega_-^2}  + \frac{1}{\omega_+^2 \omega_-^4} \bigg) \Big(
 q_1^2 l^2 ( \vec
\sigma_1 \cdot \vec \sigma_2 \, )  +  q_1^2 (\vec l \cdot \vec
\sigma_1 \, ) (\vec l \cdot \vec \sigma_2 \, )  +  (\vec q_1 \cdot
\vec l \, )  (\vec q_1 \cdot \vec
\sigma_1 \, ) (\vec l \cdot \vec \sigma_2 \, ) \nn
&& {} +  (\vec q_1 \cdot
\vec l \, )  (\vec l \cdot \vec
\sigma_1 \, ) (\vec q_1 \cdot \vec \sigma_2 \, )  -  (\vec q_1 \cdot
\vec l \, )^2  ( \vec
\sigma_1  \cdot \vec \sigma_2 \, ) -   
l^2  (\vec q_1 \cdot \vec
\sigma_1 \, ) (\vec q_1 \cdot \vec \sigma_2 \, )  \Big)
+ 6  \bigg( \frac{1}{\omega_+^4
  \omega_-^2}  + \frac{1}{\omega_+^2 \omega_-^4} \bigg) \nn 
&& {} \times \Big(
 - q_1^2 l^2 ( \vec
\sigma_1 \cdot \vec \sigma_2 \, )  +  q_1^2 (\vec l \cdot \vec
\sigma_1 \, ) (\vec l \cdot \vec \sigma_2 \, )  -  (\vec q_1 \cdot
\vec l \, )  (\vec q_1 \cdot \vec
\sigma_1 \, ) (\vec l \cdot \vec \sigma_2 \, ) 
 -  (\vec q_1 \cdot
\vec l \, )  (\vec l \cdot \vec
\sigma_1 \, ) (\vec q_1 \cdot \vec \sigma_2 \, ) \nn 
&& {} +  (\vec q_1 \cdot
\vec l \, )^2  ( \vec
\sigma_1  \cdot \vec \sigma_2 \, ) +   
l^2  (\vec q_1 \cdot \vec
\sigma_1 \, ) (\vec q_1 \cdot \vec \sigma_2 \, ) \Big)
\bigg\} \nn
&\stackrel{\rm DR}{=} & \frac{g_A^4 C_T}{ 48 \pi  F_\pi^4} 
\, \bigg\{ 2 \fet \tau_1 \cdot \fet \tau_2 \,  ( \vec \sigma_2 \cdot \vec
\sigma_3 ) \bigg[ 3 M_\pi - \frac{M_\pi^3}{4 M_\pi^2 + q_1^2} + 2 (2
M_\pi^2 + q_1^2) A(q_1) \bigg] \nn
&& {}+ 9 \Big[ (\vec q_1 \cdot \vec
\sigma_1 ) (\vec q_1 \cdot \vec \sigma_2 ) - q_1^2 (\vec \sigma_1 \cdot
\vec \sigma_2)  \Big] A (q_1) \bigg\}\,,
\eeqa
where the loop function $A(q)$ is defined according to
\beq
A(q) = \frac{1}{2q} \arctan \left( \frac{q}{2 M_\pi} \right) \,. 
\eeq 
Notice that we have exploited the fermionic nature of the nucleons in
order to simplify the above expression and made use of the following identities:
\beqa
\Big( (\fet \tau_2 + \fet  \tau_3 ) \sigma_2^i \sigma_3^j \Big)_{\rm A23}
&=& -\frac{1}{4}  (\fet \tau_2 + \fet  \tau_3 ) \, \delta_{ij} \left(
  1 - \vec \sigma_2 \cdot \vec \sigma_3 \right)  + \ldots \,, \nn 
\Big( \fet \tau_2 \cdot \fet  \tau_3 \, ( \vec \sigma_2 + \vec \sigma_3 \,)  \Big)_{\rm A23}
&=& - 3 \Big(  \vec \sigma_2 + \vec \sigma_3 \,  \Big)_{\rm A23} \nn 
&=&
 -\frac{3}{4}   \left(
  \vec \sigma_2 + \vec \sigma_3 \right)  +  \frac{3}{4} \fet \tau_2
\cdot \fet \tau_3 \,   \left(
  \vec \sigma_2 + \vec \sigma_3 \right) \,, 
\eeqa
where the ellipses in the first line refer to terms which are
antisymmetric with respect to the interchange of the indices $i,j$ and 
therefore do not contribute to the final result.  

We now turn to diagrams (8)-(12) in Fig.~\ref{fig2} whose
contributions are proportional to $g_A^2 C_{S,T}$. We find that 
graph (12) does not contain any irreducible pieces, while the
contributions of diagrams (8)-(11) reads:
\beqa
V_{\rm 2\pi-cont} &=&  \frac{g_A^2}{ 32 F_\pi^4} 
\int \frac{d^3l}{(2 \pi)^3} \, \frac{1}{\omega_+^2 \omega_-^2}
\bigg\{ i ( C_S + C_T ) \, [\fet \tau_1 \times \fet \tau_2 ] \cdot \fet \tau_3 \, 
\Big[ - (\vec q_1 \cdot \vec \sigma_2 \, ) (\vec l \cdot \vec \sigma_3
\, ) + (\vec l \cdot \vec \sigma_2 \, ) (\vec q_1 \cdot \vec \sigma_3
\, ) \Big] \nn
&& {} + 4 C_T \fet \tau_1 \cdot \fet \tau_2 \Big[
-q_1^2 (\vec \sigma_2 \cdot \vec \sigma_3 \,)
+  (\vec q_1 \cdot \vec \sigma_2 \, ) (\vec q_1 \cdot \vec \sigma_3
\, ) +  l^2 ( \vec \sigma_2  \cdot \vec \sigma_3
\, ) -  (\vec l \cdot \vec \sigma_2 \, ) (\vec l \cdot \vec \sigma_3
\, ) - 2 i [\vec q_1 \times \vec l \,  ] \cdot \vec \sigma_3 \Big] \,.
\eeqa
It is easy to see that all terms involving the imaginary unit number $i$ 
are vanishing. Using dimensional regularization and performing
antisymmetrization with respect to the nucleons 2 and 3 we arrive at
the following final result for the $g_A^2 C_{S,T}$ contribution to
$V_{\rm 2\pi-cont}$: 
\beq 
\label{2picont2}
V_{\rm 2\pi-cont}  \stackrel{\rm DR}{=}  - \frac{g_A^2 C_T}{ 24 \pi  F_\pi^4} 
\, \fet \tau_1 \cdot \fet \tau_2  \, ( \vec \sigma_2 \cdot \vec
\sigma_3 ) \Big[ M_\pi  +  (2
M_\pi^2 + q_1^2) A(q_1) \Big] \,.
\eeq  

Finally, it is easy to see that the last diagram in Fig.~\ref{fig2}
does not contribute to 3NF as the corresponding Feynman graph involves
at this order in the chiral expansion only reducible pieces. 
To summarize, the complete contribution of the
two-pion-exchange-contact topology of the 3NF at N$^3$LO 
is given by Eqs.~(\ref{2picont1}) and (\ref{2picont2}) using  the
convention of Eq.~(\ref{convention}).

\section{Leading relativistic corrections}
\def\theequation{\arabic{section}.\arabic{equation}}
\label{sec4}

The leading relativistic corrections -- that is the corrections of the
operators in $1/m$ -- to the 3NF provide further
contributions to the two-pion-exchange and one-pion-exchange-contact
topologies. They emerge from two different sources. We remind the
reader that the formally leading 3NF generated
by the tree-level two-pion-exchange \footnote{There are two
  such diagrams: the one proportional to $g_A^4$
and the one involving a Weinberg-Tomozawa vertex which is proportional
to $g_A^2$.} and one-pion-exchange contact diagrams vanishes at NLO. 
Stated differently, the resulting contributions to the scattering
amplitude are either purely reducible (at that order) or shifted to
N$^3$LO due to the suppression 
by one power of $Q/m$ caused by the
time derivative entering the Weinberg-Tomozawa vertex
\cite{Weinberg:1990rz,vanKolck:1994yi}.
The first kind of relativistic corrections emerges from taking into
account retardation effects in NLO diagrams, see graphs (1), (2) and
(6) in Fig.~\ref{fig3}. As will be shown below, contrary to the case
when the nucleons are treated as static sources, these diagrams do
produce non-vanishing 3NFs. Secondly, one also needs to take into
account the $1/m$-corrections to the leading $\pi NN$  and $\pi\pi NN$   
vertices which leads to graphs (3), (4), (5) and (7) in that
figure. Notice that there are no $1/m$-corrections to the leading
two-nucleon contact interactions.    

We begin with the retardation corrections to the two-pion exchange 3NF
corresponding to diagram (1) in Fig.~\ref{fig3}. When evaluating its
contribution to the potential, one needs, in addition to the terms listed
in Eq.~(A.1) of Ref.~\cite{Epelbaum:2007us}, to take into account effects
induced by the additional unitary transformation driven by the operator $S_8$ in
Eq.~(4.24) of Ref.~\cite{Kolling:2011mt} which is parametrized in
that work by a constant $\bar \beta_8$. The explicit form of the
operators to be evaluated is given in Eq.~(\ref{retardg4}). The resulting 3NF contribution
has the form 
\beqa
\label{rel1}
V_{2\pi , \; 1/m} &=& -\frac{g_A^4}{32 m F_\pi^4} \frac{( \vec \sigma_1
  \cdot \vec q_1 \, ) (\vec \sigma_3
  \cdot \vec q_3 \, )}{ (q_1^2 + M_\pi^2)^2(q_3^2 + M_\pi^2)}
\Big[ (1 - 2 \bar \beta_8) \Big( \fet \tau_1 \cdot \fet \tau_3 \,
(\vec q_1 \cdot \vec q_3  )^2 + [\fet \tau_1 \times \fet \tau_2]
\cdot \fet \tau_3 \, [\vec q_1 \times \vec q_3] \cdot \vec \sigma_2 \,
(\vec q_1 \cdot \vec q_3  ) \Big) \nn
&-&  2 i \Big( \fet\tau_1 \cdot \fet \tau_3 \, [
  \vec q_1 \times \vec q_3 ] \cdot \vec \sigma_2  -  [\fet \tau_1 \times \fet \tau_2]
\cdot \fet \tau_3 \,(\vec q_1 \cdot \vec q_3  ) \Big)  \left(  (1 - 2
  \bar \beta_8) \, (\vec q_1
    \cdot \vec k_2)
+  (1 + 2 \bar \beta_8) \, ( \vec q_1
    \cdot \vec k_1) \right) \Big]\,,
\eeqa
where $\vec k_i \equiv (\vec p_i + \vec p_i \,  ' )/2$.  Notice that the
unitary transformation considered above also affects the form of
the $1/m^2$-corrections to the one-pion exchange and $1/m$-corrections
to the two-pion exchange two-nucleon potentials at N$^3$LO. To be
consistent with the N$^3$LO potential of 
Ref.~\cite{Epelbaum:2004fk}, one has to choose $\bar \beta_8 = 1/4$.  
The contribution of diagram (2) in  Fig.~\ref{fig3} is not
affected by the above UT and has the form
\beq
\label{rel2}
V_{2\pi , \; 1/m} = i \frac{g_A^2}{32 m F_\pi^4} \, \frac{( \vec \sigma_1
  \cdot \vec q_1 \, )(\vec \sigma_3
  \cdot \vec q_3 \, )}{ (q_1^2 + M_\pi^2)(q_3^2 + M_\pi^2)}
\, [\fet \tau_1 \times \fet \tau_2]
\cdot \fet \tau_3 \, (\vec q_3 \cdot \vec k_3 - \vec q_1 \cdot \vec
k_1 ) \,.
\eeq
Finally, retardation corrections to the one-pion-exchange-contact
topology from diagram (6) in Fig.~\ref{fig3} have the form
\beqa
\label{rel3}
V_{1\pi{\rm -cont}, \; 1/m} &=& \frac{g_A^2}{8 m F_\pi^2} \frac{\vec \sigma_1
  \cdot \vec q_1 }{ (q_1^2 + M_\pi^2)^2}\, \fet \tau_1 \cdot \fet
\tau_2 \, \Big\{ (1 - 2 \bar \beta_8) \,  (\vec q_1 \cdot \vec q_3 )  
\, \left[ C_S (\vec q_1 \cdot \vec \sigma_2 ) + C_T (\vec q_1 \cdot
 \vec  \sigma_3  ) \right]  + 2 i C_T 
  \vec q_1 \cdot [ \vec \sigma_2 \times \vec \sigma_3 ] \nn
&& {} \times \left[  (1 - 2 \bar \beta_8) \, (\vec q_1\cdot \vec k_2 )
  +  (1 + 2 \bar \beta_8) \, (\vec q_1\cdot \vec k_1 ) \right]\Big\}\,.
\eeqa

We now turn to contributions emerging from relativistic corrections to
vertices in the pion-nucleon Hamilton density. Consider first the
contributions of  diagrams
(3), (4) and (7) in Fig.~\ref{fig3}.  The relevant terms in
the effective Lagrangian/Hamiltonian have the form\footnote{Different
  conventions for the sign of the axial vector coupling constant $g_A$
are used in the literature. For example, $g_A$ in  Ref.~\cite{Epelbaum:2005fd}
corresponds to $-g_A$ in the present work. This difference does not
affect the expressions for nuclear forces and currents where the LEC $g_A$
enters quadratically. It is important, however, to use the same
convention for both terms in Eq.~(\ref{lagr}).   }
\beq
\label{lagr}
\mathcal{L}_{\pi N} = - \mathcal{H}_{\pi N} = - \frac{g_A}{2 F_\pi}
N^\dagger \fet \tau \vec \sigma \cdot \vec \nabla \fet \pi N - 
i \frac{g_A}{4 m F_\pi} N^\dagger \fet \tau \vec \sigma  \cdot \left(
  \overleftarrow{\nabla} -    \overrightarrow{\nabla} \right) N \cdot
\dot{\fet \pi}\,,
\eeq
where $N$ and $\fet \pi$ refer to the nucleon and pion fields,
respectively. Similarly to the retardation corrections considered
above, it is possible to construct an additional UT with the
generator proportional to $g_A^2/m$ which affects the contributions of
diagrams (3) and (7) when acting on the one-pion exchange and the
leading contact potentials, respectively. We adopt here the notation
of Ref.~\cite{Kolling:2011mt}, where this unitary operator is
parameterized in terms of another arbitrary parameter $\bar \beta_9$,
see Eq.~(4.24) of that work. Again, the UT also affects the
$1/m^2$- and $1/m$-corrections to the one-pion and two-pion exchange potential
at N$^3$LO, respectively, so that the parameter $\bar \beta_9$ needs to be chosen
consistently. Using the explicit expressions for the operator
structure of the unitarily transformed Hamilton operator in Eq.~(A.4)
of Ref.~\cite{Epelbaum:2007us} and adding terms induced by the above
mentioned UT we obtain the following 3NF contribution from diagram
(3):
\beqa 
\label{rel4}
V_{2\pi , \; 1/m} &=& - \frac{g_A^4}{32 m F_\pi^4} \, \frac{ \vec \sigma_1
  \cdot \vec q_1 \, }{ (q_1^2 + M_\pi^2)(q_3^2 + M_\pi^2)}
\, \bigg\{ \fet \tau_1 \cdot \fet \tau_3 \, \Big[ (2 \bar \beta_9 - 1
  ) \, (\vec \sigma_3 \cdot \vec q_3 ) \, \left( q_1^2 + 2 i [\vec q_1
    \times \vec k_2 ] \cdot \vec \sigma_2 \right)  - 2 i  (2 \bar \beta_9 + 1
  ) \, (\vec \sigma_3 \cdot \vec k_3 ) \nn
&& {} \times   [\vec q_1
    \times \vec q_3 ] \cdot \vec \sigma_2 \Big]  + 2 i [\fet \tau_1
    \times \fet \tau_2 ] \cdot \fet \tau_3 \,  \Big[ - (2 \bar \beta_9 - 1
  ) \, (\vec \sigma_3 \cdot \vec q_3 ) \,   (\vec q_1 \cdot \vec
  k_2 )   +   (2 \bar \beta_9 + 1
  ) \, (\vec \sigma_3 \cdot \vec k_3 )  (\vec q_1 \cdot \vec
  q_3 )   \Big] \bigg\}\,.
\eeqa
Next, it is easy to see that diagram (4) does not generate any
3NF at the considered order in the chiral expansion. Similarly to the
corresponding NLO diagram (same topology constructed from the
lowest-order vertices), there is suppression by, in this case, two
factors of $Q/m$ due to the time derivatives entering the
Weinberg-Tomozawa vertex and the $1/m$-correction to the $\pi NN$
vertex in Eq.~(\ref{lagr}). Consequently, within the power counting
scheme adopted in the present work, this diagram contributes to the 3NF
only at order N$^5$LO.     
Finally, the contribution of diagram (7) in Fig.~\ref{fig3} is clearly also
affected by the abovementioned unitary transformation and is given by 
\beqa
\label{rel5}
V_{1\pi{\rm -cont}, \; 1/m} &=&-\frac{g_A^2 C_T}{8 m F_\pi^2} \frac{1 }{q_1^2 
+ M_\pi^2}\, \fet \tau_1 \cdot \fet
\tau_2 \, \Big\{ 2 i (2 \bar \beta_9 + 1) \,  (\vec k_1 \cdot \vec \sigma_1 )  
\, \vec q_1 \cdot [ \vec \sigma_2 \times \vec \sigma_3 ] 
- (2 \bar \beta_9 - 1) \,  (\vec q_1 \cdot \vec \sigma_1 )  \,  (\vec
q_3 \cdot \vec \sigma_3 )  \\
&& {} - 2 i (2 \bar \beta_9 - 1) \,  (\vec q_1 \cdot \vec \sigma_1 )  
\, \vec k_2 \cdot [ \vec \sigma_2 \times \vec \sigma_3 ] \Big\}  
+ \frac{g_A^2 C_S}{8 m F_\pi^2} \frac{1 }{ q_1^2 + M_\pi^2}\, \fet \tau_1 \cdot \fet
\tau_2 \,   (2 \bar \beta_9 - 1) \,  (\vec q_1 \cdot \vec \sigma_1 )  \,  (\vec
q_3 \cdot \vec \sigma_2 ) \,.
\nonumber
\eeqa

Finally, consider the contribution of diagram (5). Since this graph
does not induce any reducible pieces, the corresponding 3NF can be
identified with the amplitude and computed using the Feynman-diagram
technique.  The Feynman rule for the $1/m$-correction to the $\pi\pi
NN$ vertex can be found e.g.~in Ref.\cite{Bernard:1995dp}. In our
notation the relevant terms\footnote{We do not show terms involving
  zeroth components of the nucleon momenta since they do not
  contribute to the 3NF at N$^3$LO.}  have the form
\beq
 - \frac{1}{8 m F_\pi^2} \epsilon^{abc} \tau^c (\vec p + \vec p \, ')
 \cdot (\vec q_1 + \vec q_2 ) -   \frac{i}{4 m F_\pi^2} \epsilon^{abc}
 \tau^c [\vec q_1 \times    \vec q_2 ] \cdot \vec \sigma \,,
\eeq 
where $\vec q_1$, $a$ ($\vec q_2$, $b$) denote the momentum and isospin quantum number
of the incoming (outgoing) pion.  We then obtain for the contribution of
diagram (5):
\beq
\label{rel6}
V_{2\pi , \; 1/m} = \frac{g_A^2}{32 m F_\pi^4} \, \frac{ (\vec \sigma_1
  \cdot \vec q_1 )( \vec \sigma_3 \cdot \vec q_3 ) }{ (q_1^2 + M_\pi^2)(q_3^2 + M_\pi^2)}
\, [ \fet \tau_1 \times \fet \tau_3 ] \cdot \fet \tau_2 \, \left(
  [\vec q_1 \times \vec q_3] \cdot \vec \sigma_2 + i \vec k_2 \cdot
  (\vec q_3 - \vec q_1) \right)\,.
\eeq

To summarize, our results for the relativistic ($1/m$) corrections to the 3NF
at N$^3$LO are given by Eqs.~(\ref{rel1}), (\ref{rel2}), (\ref{rel3}),
(\ref{rel4}), (\ref{rel5}) and (\ref{rel6}). Adding up these
expressions, the two-pion exchange
contribution can be written in the form 
\beq
\label{2pi_rel_mom}
V_{2\pi , \; 1/m} = \frac{g_A^2}{32 m F_\pi^4} \, \frac{1}{ (q_1^2 +
  M_\pi^2)(q_3^2 + M_\pi^2)}
\, \Big( \fet \tau_1 \cdot \fet \tau_3 \, A_{123} + [ \fet \tau_1 \times
\fet \tau_2] \cdot \fet \tau_3 \, B_{123} \Big)\; +\;\mbox{5
  permutations} \,, 
\eeq
where 
\beqa
A_{123} &=& ( \vec \sigma_1 \cdot \vec q_1  ) ( \vec \sigma_3 \cdot \vec
q_3  ) \, a_{123} + ( \vec \sigma_1 \cdot \vec q_1  ) ( \vec \sigma_3 \cdot \vec
k_3  )  \, c_{123}\,, \nn
B_{123} &=& ( \vec \sigma_1 \cdot \vec q_1  ) ( \vec \sigma_3 \cdot \vec
q_3  ) \, b_{123} + ( \vec \sigma_1 \cdot \vec q_1  ) ( \vec \sigma_3 \cdot \vec
k_3  )  \, d_{123}\,,
\eeqa
and the functions $a_{123}$,   $b_{123}$,   $c_{123}$ and   $d_{123}$
are given by:  
\beqa
\label{abcd}
a_{123}&=& - \frac{g_A^2}{q_1^2 + M_\pi^2} \Big[ ( 1 - 2 \bar \beta_8)
(\vec q_1 \cdot \vec q_3)^2 - 2 i [ \vec q_1 \times \vec q_3] \cdot
\vec \sigma_2  \left( (1 - 2 \bar \beta_8 ) ( \vec q_1 \cdot \vec k_2
  )
+ ( 1 + 2 \bar \beta_8 ) (\vec q_1 \cdot \vec k_1 ) \right) \Big]  \nn
&& {} - g_A^2 (2 \bar \beta_9 - 1 ) \left( q_1^2 + 2 i [\vec q_1
  \times \vec k_2] \cdot \vec \sigma_2  \right)\,, \nn
c_{123} &=& 2 i g_A^2 (2 \bar \beta_9 + 1) [ \vec q_1 \times \vec q_3]
\cdot \vec \sigma_2 \,, \nn
b_{123} &=&  - \frac{g_A^2}{q_1^2 + M_\pi^2} \Big[ ( 1 - 2 \bar
\beta_8)\, 
[\vec q_1 \times \vec q_3 ] \cdot \vec \sigma_2 \, (\vec q_1 \cdot
\vec q_3 ) + 2 i (\vec q_1 \cdot \vec q_3 ) \left( (1 - 2 \bar \beta_8
  ) (\vec q_1 \cdot \vec k_2 ) + (1 + 2 \bar \beta_8 ) (\vec q_1 \cdot
  \vec k_1) \right) \Big] \nn
&& {} + 2 i \vec q_3 \cdot (\vec k_3 - \vec k_2 ) + 2 i g_A^2 (2 \bar
\beta_9 - 1 ) (\vec q_1 \cdot \vec k_2) - [\vec q_1 \times \vec q_3 ]
\cdot \vec \sigma_2 \,, \nn
d_{123} &=& - 2 i g_A^2 (2 \bar \beta_9 + 1) (\vec q_1 \cdot \vec q_3)\,.
\eeqa
The relativistic corrections to the one-pion-exchange-contact topology
have the form 
\beq
\label{1pi_cont_rel_mom}
V_{1\pi{\rm -cont}, \; 1/m} = \frac{g_A^2}{8 m F_\pi^2} \, \frac{1}{ q_1^2 +
  M_\pi^2} \, \fet \tau_1 \cdot \fet \tau_2 
\, \Big( (\vec \sigma_1 \cdot \vec q_1 ) \, f_{123} + (\vec \sigma_1
\cdot \vec k_1 ) \, g_{123}  \Big)\; +\;\mbox{5
  permutations} \,, 
\eeq
where 
\beqa
f_{123} &=& \frac{1}{q_1^2 + M_\pi^2} \Big[ 
(1 - 2 \bar \beta_8 ) (\vec q_1 \cdot \vec q_3 )
\Big( C_S (\vec q_1 \cdot \vec \sigma_2 ) + C_T (\vec q_1 \cdot \vec
  \sigma_3 ) \Big) + 2 i C_T \vec q_1 \cdot [ \vec \sigma_2 \times
\vec \sigma_3 ] \Big( (1 - 2 \bar \beta_8 ) (\vec q_1 \cdot \vec k_2 )
+ (1 + 2 \bar \beta_8) \nn
&& {} \times (\vec q_1 \cdot \vec k_1 ) \Big) \Big] 
+ (2 \bar \beta_9 - 1 ) \Big[ C_S (\vec q_3 \cdot \vec \sigma_2 ) + C_T (\vec q_3 \cdot \vec
  \sigma_3 ) \Big] + 2 i C_T (2 \bar \beta_9 - 1 ) \vec k_2 \cdot [
  \vec \sigma_2 \times \vec \sigma_3 ]\,, \nn
g_{123} &=& - 2 i C_T (2 \bar \beta_9 + 1 ) \vec q_1 \cdot [\vec
\sigma_2 \times \vec \sigma_3 ]    \,.
\eeqa
The above expressions are
written in a general form including the dependence on the constants $\bar
\beta_{8,9}$ which parametrize the unitary ambiguity of these
potentials. The $1/m^2$-corrections ($1/m$-corrections) to the
one-pion exchange (two-pion-exchange)  two-nucleon potential at
N$^3$LO are also
affected by this unitary ambiguity. A comprehensive
discussion on this issue and on the relation between different forms of relativistic
extensions of the Schr\"odinger equation and non-static terms in the
two-nucleon potential can be found in Ref.~\cite{Friar:1999sj}.  To be consistent with the two-nucleon potential of
Ref.~\cite{Epelbaum:2004fk}, one needs to choose 
\beq
\bar \beta_8 = \frac{1}{4}\,, \quad \quad \bar \beta_9 = 0\,.
\eeq
We emphasize that in order to completely take into account the 
relativistic corrections at N$^3$LO in few-body calculations, one needs to
use the Schr\"odinger equation with the relativistic expression for the nucleon kinetic
energy, the proper forms of the $1/m^2$-corrections to the
$1\pi$-exchange NN potential as well as $1/m$-corrections to the   
NN and 3N forces. In addition, one has to take into account boost
corrections to the leading-order two-nucleon potential (i.e.~the $\vec
P$-dependence in the $1\pi$-exchange potential and the leading contact
interactions, see Ref.~\cite{Girlanda:2010ya}).  

While  $1/m$-corrections to the
one-pion-exchange-contact 3NF given in Eq.~(\ref{1pi_cont_rel_mom}) have been
never considered before, the leading relativistic corrections to the
two-pion exchange 3NF were studied  by Friar and Coon in
Ref.~\cite{Friar:1994zz}, who, however, only provide expressions in
coordinate space. 
In order to facilitate a comparison between our results and the ones
by Friar and Coon, we give in appendix \ref{app2} the coordinate-space
representation of the obtained 3NF contributions. 
Our profile functions are related to the one used in
Ref.~\cite{Friar:1994zz}  according to
\beq
U_1 (M_\pi r ) = \frac{4 \pi }{M_\pi} h_0 (r)\,, \quad \quad
\vec \nabla_{\vec r} \, U_2 (M_\pi r ) = - 4 \pi M_\pi \vec r \, h_0
(r)\,,
\eeq
where the function $h_0$ is defined in Eq.~(24c) of
that work.  Further, $f^2$ ($M$) in Ref.~\cite{Friar:1994zz} correspond
to $(g_A/(2 F_\pi))^2$ ($m$) in our notation. Friar and Coon also
discuss the unitary ambiguity of the resulting 3N potentials
associated with the UTs $\propto 1/m$. They parametrized it in terms of
the arbitrary parameters $\mu$ and $\nu$, see also
Ref.~\cite{Friar:1999sj}. 
The relations between their $\mu$, $\nu$ and our
$\bar \beta_{8,9}$ are given by
\beq
\mu = 4 \bar \beta_9 + 1 \,, \quad \quad 
\nu = 2 \bar \beta_8\,.
\eeq     
Taking into account the typographical error in \cite{Friar:1994zz} as
pointed out in [5] of Ref.~\cite{Friar:1999sj}, we observe that our
results for the $1/m$-corrections to the two-pion exchange 3NF in
Eqs.(\ref{2pi_rel_coord})-(\ref{rspace}) agree with the ones in
Eqs.~(23a), (23b), (23c), (33a), (33b), (33c) and (33d) of
Ref.~\cite{Friar:1994zz} except for the
retardation corrections in $\tilde a_{123}$ and $\tilde b_{123}$
emerging from diagram (1) in Fig.~\ref{fig3}.  More precisely, we
reproduce the results of Friar and Coon if we make the replacement 
\beq
  \left( (1 - 2 \bar \beta_8 ) ( \vec q_1 \cdot \vec k_2 )
+ ( 1 + 2 \bar \beta_8 ) (\vec q_1 \cdot \vec k_1 ) \right)
\; \longrightarrow \; 
\left( (1 - 2 \bar \beta_8 ) ( \vec q_1 \cdot \vec k_2 )
+ ( 1 + 2 \bar \beta_8 ) (\vec q_1 \cdot \vec k_1 ) \right) + \delta
\eeq
in $a_{123}$ and $b_{123}$ in Eq.~(\ref{abcd}) with $\delta$ given by 
\beq
\delta =   (1 - 2 \bar \beta_8 ) \,  \vec q_1 \cdot ( \vec k_2 - \vec
k_1 ) +  (1 - 2 \bar \beta_8 ) \,  \vec q_3 \cdot ( \vec k_2 - \vec
k_3 ) = \frac{1}{2} \, (1 - 2 \bar \beta_8 ) \, \left( p_1^2 + p_2^2 +
  p_3^2 - {p_1'}^2  - {p_2'}^2  - {p_3'}^2 \right)\,,
\eeq
where we made use of the identity $\vec q_1 + \vec q_2 + \vec q_3 =
0$. This implies that the difference between our results
appears only when the nucleons are off-the-energy shell.  We
conjecture that this difference originates from the additional UTs $\exp
(\alpha_1 S_1 + \alpha_2 S_2 )$ with $\alpha_{1,2}$ denoting the
transformation angles and $S_{1,2}$ the generators of the   
two-pion-exchange range whose explicit form is given in Eq.~(3.25) of
Ref.~\cite{Epelbaum:2007us}. 
The corresponding unitary ambiguity is \emph{not} explored by Friar
and Coon. It is easy to see that these UTs
induce, among others,  terms in the 3NF $\propto [ E_{\rm kin}, \;
(\alpha_1 S_1 + \alpha_2 S_2)]$ which have the same  structure as
the $\delta$-terms discussed above.  As explained in
Ref.~\cite{Epelbaum:2007us}, the angles $\alpha_{1,2}$ have to be
chosen in a specific way, see Eq.~(3.31) of that work, in order to
maintain renormalizability of the leading loop corrections to the
3NF. Our findings therefore indicate that the particular choice made
in Ref.~\cite{Friar:1994zz}  for $\alpha_{1,2}$ is \emph{not} compatible with the abovementioned
renormalizability constraint.

\section{Summary and conclusions}
\def\theequation{\arabic{section}.\arabic{equation}}
\label{sec5}

In this work, we have derived the short-range part and the relativistic
($1/m$) corrections to the 3NF in chiral effective field theory at N$^3$LO.
Combined with the long-range parts already given in
Ref.~\cite{Bernard:2007sp}, this completes the calculation of two-, three-
and four-nucleon forces at this order in the formulation without
explicit $\Delta$(1232) degrees of freedom.  The short-range parts considered
here consists formally of two different topologies. Remarkably, there is no contribution
from the one-pion-exchange-contact topology at N$^3$LO as we have demonstrated
in Sec.~\ref{sec2}. The total contribution from the two-pion-exchange-contact topology 
is given in Eqs.~(\ref{2picont1}), (\ref{2picont2}) supplemented with
Eq.~(\ref{convention}). 
Further, there are relativistic
($1/m$) corrections to the leading 3NF at N$^2$LO, which are worked out in
Sec.~\ref{sec4}. The corresponding terms of the 3NF are summarized in
Eqs.~(\ref{2pi_rel_mom}) and (\ref{1pi_cont_rel_mom}).  We also
provide the coordinate-space representation of the obtained results  in appendix
\ref{app2} and compare our findings for the long-range
$1/m$-corrections with the earlier calculation by Friar and Coon
\cite{Friar:1994zz}. 
  
We stress again that at N$^3$LO, the 3NF is free of unknown LECs, so the complete
3NF consisting of terms at N$^2$LO and N$^3$LO contains altogether only two
LECs $D$ and $E$ which need to be determined from few-nucleon data. 
With the results presented here, one is now in the position to analyze in
detail the many data in few-nucleon and heavier systems based on consistent
and precise two- and three-nucleon forces. In particular, it will be
interesting to find out whether the remaining discrepancies in the three- 
and four-nucleon systems will be overcome employing the force derived in
Ref.~\cite{Bernard:2007sp} and here. Work along these lines is in progress.

\section*{Acknowledgments}

This work was
supported by funds provided by the Helmholtz Association (grants VH-NG-222 and
VH-VI-231), by the DFG (SFB/TR 16 ``Subnuclear
Structure of Matter''), by the EU HadronPhysics2 project ``Study
of strongly interacting matter'' and the European Research Council
(ERC-2010-StG 259218 NuclearEFT) and by BMBF (grant 06BN9006).

\appendix

\def\theequation{\Alph{section}.\arabic{equation}}
\setcounter{equation}{0}
\section{Formal algebraic structure of the 3NF corrections}
\label{app1}

In this appendix we list the formal
operator structure of the various N$^3$LO contributions to the nuclear
Hamiltonian relevant for the present calculations. A detailed
discussion on these terms can be found in Ref.~\cite{Epelbaum:2007us}.
We include contributions induced by the additional UTs considered in
that work and in Eq.~(4.23) of Ref.~\cite{Kolling:2011mt}. Except
for $\alpha_5$ and $\bar \beta_{8,9}$,  the corresponding 
transformation angles are fixed by the
renormalizability constraints as 
discussed in \cite{Epelbaum:2007us}. In all applications to nuclear
forces and currents considered so far, the dependence on $\alpha_5$
drops in the final results.   
\begin{itemize}
\item{terms $\propto g_A^4 C_{S,T}$:}
\beqa
\label{class4}
V &=&  \eta \bigg[  - \frac{1}{2}  
  H_{21}^{(1)} \frac{\lambda^1}{E_\pi} H_{21}^{(1)} \, \eta \, H_{21}^{(1)} \frac{\lambda^1}{E_\pi} H_{40}^{(2)}
  \frac{\lambda^1}{E_\pi^2} H_{21}^{(1)}  
 \; - \; \frac{1}{2}  
  H_{21}^{(1)} \frac{\lambda^1}{E_\pi} H_{21}^{(1)} \, \eta \, H_{21}^{(1)} \frac{\lambda^1}{E_\pi^2} H_{40}^{(2)}
  \frac{\lambda^1}{E_\pi} H_{21}^{(1)}   
  \nn [2pt]
&& {}
 - \frac{1}{2}  
  H_{21}^{(1)} \frac{\lambda^1}{E_\pi^2} H_{21}^{(1)} \, \eta \, H_{21}^{(1)} \frac{\lambda^1}{E_\pi} H_{40}^{(2)}
  \frac{\lambda^1}{E_\pi} H_{21}^{(1)}  
\; - \;  \frac{1}{2}  
  H_{40}^{(2)} \, \eta \,  H_{21}^{(1)} \frac{\lambda^1}{E_\pi} H_{21}^{(1)}
  \frac{\lambda^2}{E_\pi} H_{21}^{(1)}
  \frac{\lambda^1}{E_\pi^2} H_{21}^{(1)}   \nn [2pt]
&& {}
- \frac{1}{2}  
  H_{40}^{(2)} \, \eta \,  H_{21}^{(1)} \frac{\lambda^1}{E_\pi} H_{21}^{(1)}
  \frac{\lambda^2}{ E_\pi^2} H_{21}^{(1)}
  \frac{\lambda^1}{E_\pi} H_{21}^{(1)}   
\; - \;  \frac{1}{2}  
  H_{40}^{(2)} \, \eta \,  H_{21}^{(1)} \frac{\lambda^1}{E_\pi^2} H_{21}^{(1)}
  \frac{\lambda^2}{E_\pi} H_{21}^{(1)}
  \frac{\lambda^1}{E_\pi} H_{21}^{(1)}    \nn [2pt]
&& {}
+ \frac{1}{2}  
  H_{40}^{(2)} \, \eta \,  H_{21}^{(1)} \frac{\lambda^1}{E_\pi} H_{21}^{(1)}
  \, \eta \,  H_{21}^{(1)}
  \frac{\lambda^1}{E_\pi^3} H_{21}^{(1)}   
\; + \;  \frac{3}{8}  
  H_{40}^{(2)} \, \eta \,  H_{21}^{(1)} \frac{\lambda^1}{E_\pi^2} H_{21}^{(1)}
  \, \eta \,  H_{21}^{(1)}
  \frac{\lambda^1}{E_\pi^2} H_{21}^{(1)}  \nn [2pt]
&& {}
  +   \frac{1}{8}  
  H_{21}^{(1)} \frac{\lambda^1}{E_\pi^2} H_{21}^{(1)} \, \eta \, H_{40}^{(2)} \, \eta \,  H_{21}^{(1)}
  \frac{\lambda^1}{E_\pi^2} H_{21}^{(1)}   
 \; + \;   
  H_{21}^{(1)} \frac{\lambda^1}{E_\pi} H_{21}^{(1)} \frac{\lambda^2}{E_\pi}
  H_{21}^{(1)}  \frac{\lambda^1}{E_\pi} H_{40}^{(2)} \frac{\lambda^1}{E_\pi} H_{21}^{(1)}  \nn[2pt]
&& {}  +   \frac{1}{2} 
  H_{21}^{(1)} \frac{\lambda^1}{E_\pi} H_{21}^{(1)} \frac{\lambda^2}{E_\pi}
  H_{40}^{(2)}  \frac{\lambda^2}{E_\pi} H_{21}^{(1)} \frac{\lambda^1}{E_\pi} H_{21}^{(1)}  
\; + \;   H_{21}^{(1)} \frac{\lambda^1}{E_\pi}  H_{21}^{(1)} \,
\eta \, H_{21}^{(1)} \frac{\lambda^1}{E_\pi^3}  H_{21}^{(1)} \, \eta \,
H_{40}^{(2)} \nn [2pt]
&& {}
-  \frac{1}{2} \, H_{21}^{(1)} \frac{\lambda^1}{E_\pi^3}  H_{21}^{(1)} \,
\eta \, H_{21}^{(1)} \frac{\lambda^1}{E_\pi}  H_{21}^{(1)} \, \eta \, H_{40}^{(2)}  
\; - \; \frac{1}{4} \, H_{21}^{(1)} \frac{\lambda^1}{E_\pi}  H_{21}^{(1)}  \frac{\lambda^2}{E_\pi}
 H_{21}^{(1)} \frac{\lambda^1}{E_\pi^2}  H_{21}^{(1)} \, \eta \, H_{40}^{(2)} \nn [2pt]
&& {} + \frac{1}{4} \, H_{21}^{(1)} \frac{\lambda^1}{E_\pi^2}  H_{21}^{(1)}  \frac{\lambda^2}{E_\pi}
 H_{21}^{(1)} \frac{\lambda^1}{E_\pi}  H_{21}^{(1)} \, \eta \, H_{40}^{(2)} \bigg] \eta + \mbox{h.c.} 
\eeqa
\item{terms $\propto g_A^2 C_{S,T}$:}
\beqa
\label{class5}
V&=&  \eta \bigg[  -2 \alpha_5  
  H_{40}^{(2)} \, \eta \, H_{21}^{(1)} \frac{\lambda^1}{E_\pi} H_{21}^{(1)} \frac{\lambda^2}{E_\pi^2} H_{22}^{(2)}
\; + \;  
\left( \frac{1}{2} + \alpha_5 \right) 
  H_{40}^{(2)} \, \eta \, H_{21}^{(1)} \frac{\lambda^1}{E_\pi} H_{22}^{(2)}
  \frac{\lambda^1}{E_\pi^2} H_{21}^{(1)} \nn [2pt]
&& {}  +  
\left( \frac{1}{2} + \alpha_5 \right)   
  H_{40}^{(2)} \, \eta \, H_{22}^{(2)} \frac{\lambda^2}{E_\pi} H_{21}^{(1)}
  \frac{\lambda^1}{E_\pi^2} H_{21}^{(1)} 
\;  + \; \left( \frac{1}{2} - \alpha_5 \right)  H_{40}^{(2)} \, \eta \,
H_{21}^{(1)} \frac{\lambda^1}{E_\pi^2} H_{21}^{(1)} \frac{\lambda^2}{E_\pi}
H_{22}^{(2)} \nn [2pt]
&& {}  +   
\left( \frac{1}{2} - \alpha_5 \right)   
  H_{40}^{(2)} \, \eta \, H_{21}^{(1)} \frac{\lambda^1}{E_\pi^2} H_{22}^{(2)} \frac{\lambda^1}{E_\pi} H_{21}^{(1)}
\; + \;  
\left( 1 + 2 \alpha_5 \right)    
  H_{40}^{(2)} \, \eta \, H_{22}^{(2)} \frac{\lambda^2}{E_\pi^2} H_{21}^{(1)}
  \frac{\lambda^1}{E_\pi} H_{21}^{(1)} \\
&& {} 
 - H_{21}^{(1)} \frac{\lambda^1}{E_\pi}  H_{21}^{(1)} \frac{\lambda^2}{E_\pi} H_{40}^{(2)} \frac{\lambda^2}{E_\pi} H_{22}^{(2)}
 - H_{21}^{(1)} \frac{\lambda^1}{E_\pi}  H_{22}^{(2)} \frac{\lambda^1}{E_\pi} H_{40}^{(2)} \frac{\lambda^1}{E_\pi} H_{21}^{(1)}
 - H_{21}^{(1)} \frac{\lambda^1}{E_\pi}  H_{40}^{(2)} \frac{\lambda^1}{E_\pi} H_{21}^{(1)} \frac{\lambda^2}{E_\pi} H_{22}^{(2)}
\bigg ] \eta + \mbox{h.c.}\,. \nonumber
\eeqa
\item{retardation corrections $\propto g_A^4/m$:}
\beqa
\label{retardg4}
V &=& \eta \bigg[\left( 
\frac{1}{2} + \bar \beta_8 \right) \,  H_{21}^{(1)} \frac{\lambda^1}{E_\pi}  H_{21}^{(1)} \, \eta \, 
      H_{21}^{(1)} \frac{\lambda^1}{E_\pi^3}  H_{21}^{(1)} \, \eta \, H_{20}^{(2)} 
\; -\;   H_{21}^{(1)} \frac{\lambda^1}{E_\pi}  H_{21}^{(1)} \, \eta \, 
      H_{21}^{(1)} \frac{\lambda^1}{E_\pi^3}  H_{20}^{(2)} \, \lambda^1 \,  H_{21}^{(1)} \nn
&& {} + \; \left( \frac{1}{2} - \bar \beta_8 \right) \,  H_{21}^{(1)} \frac{\lambda^1}{E_\pi}  H_{21}^{(1)} \, \eta \, 
      H_{20}^{(2)} \, \eta \, H_{21}^{(1)} \frac{\lambda^1}{E_\pi^3}  H_{21}^{(1)} 
\; - \; \frac{3}{4} \,  H_{21}^{(1)} \frac{\lambda^1}{E_\pi}  H_{21}^{(1)} \frac{\lambda^2}{E_\pi}  
      H_{21}^{(1)} \frac{\lambda^1}{E_\pi^2}  H_{21}^{(1)} \, \eta \,  H_{20}^{(2)} \nn
&& {} + \; H_{21}^{(1)} \frac{\lambda^1}{E_\pi}  H_{21}^{(1)} \frac{\lambda^2}{E_\pi}  
      H_{21}^{(1)} \frac{\lambda^1}{E_\pi^2}  H_{20}^{(2)} \,  \lambda^1 \, H_{21}^{(1)} 
\;  - \; \frac{1}{2} \,  H_{21}^{(1)} \frac{\lambda^1}{E_\pi}  H_{21}^{(1)} \frac{\lambda^2}{E_\pi^2}  
      H_{21}^{(1)} \frac{\lambda^1}{E_\pi}  H_{21}^{(1)} \, \eta \,  H_{20}^{(2)} \nn
&& {} + \; \frac{1}{2} \, H_{21}^{(1)} \frac{\lambda^1}{E_\pi}  H_{21}^{(1)} \frac{\lambda^2}{E_\pi^2}  
      H_{20}^{(2)} \, \lambda^2 \, H_{21}^{(1)} \frac{\lambda^1}{E_\pi}  H_{21}^{(1)} 
\; + \; \frac{3}{8} \,  H_{21}^{(1)} \frac{\lambda^1}{E_\pi^2}  H_{21}^{(1)} \, \eta \,   
      H_{21}^{(1)} \frac{\lambda^1}{E_\pi^2}  H_{21}^{(1)} \, \eta \,  H_{20}^{(2)} \nn
&& {} - \frac{1}{2} \, H_{21}^{(1)} \frac{\lambda^1}{E_\pi^2}  H_{21}^{(1)} \, \eta \, 
      H_{21}^{(1)} \frac{\lambda^1}{E_\pi^2}  H_{20}^{(2)} \, \lambda^1 \,  H_{21}^{(1)} 
\; + \; \frac{1}{8} \,  H_{21}^{(1)} \frac{\lambda^1}{E_\pi^2}  H_{21}^{(1)} \, \eta \, 
      H_{20}^{(2)} \, \eta \, H_{21}^{(1)} \frac{\lambda^1}{E_\pi^2}  H_{21}^{(1)} \nn
&& {} - \frac{1}{4} \, H_{21}^{(1)} \frac{\lambda^1}{E_\pi^2}  H_{21}^{(1)} \frac{\lambda^2}{E_\pi} 
      H_{21}^{(1)} \frac{\lambda^1}{E_\pi^2}  H_{21}^{(1)} \,\eta \,   H_{20}^{(2)}
\bigg] \eta +  \mbox{h.c.} \,,
\eeqa
\item{retardation corrections $\propto g_A^2/m$:}
\beqa
V &=& \eta \bigg[ (1 + 2 \alpha_5 ) \, 
      H_{21}^{(1)} \frac{\lambda^1}{E_\pi}  H_{21}^{(1)} \frac{\lambda^2}{E_\pi^2}
      H_{22}^{(2)} \, \eta \, H_{20}^{(2)}
\; - \;  H_{21}^{(1)} \frac{\lambda^1}{E_\pi}  H_{21}^{(1)} \frac{\lambda^2}{E_\pi^2} 
      H_{20}^{(2)} \, \lambda^2 \, H_{22}^{(2)}  \nn
&& {}
+  \frac{1}{2} (1 - 2 \alpha_5 ) \, H_{21}^{(1)} \frac{\lambda^1}{E_\pi}
      H_{22}^{(2)} \frac{\lambda^1}{E_\pi^2}  H_{21}^{(1)} \, \eta \, H_{20}^{(2)}
\; - \; H_{21}^{(1)} \frac{\lambda^1}{E_\pi} H_{22}^{(2)} \frac{\lambda^1}{E_\pi^2}
H_{20}^{(2)} \, \lambda^1 \, H_{21}^{(1)}  \nn
&& {}
+  \frac{1}{2} (1 + 2 \alpha_5 ) \, H_{21}^{(1)} \frac{\lambda^1}{E_\pi^2}  H_{21}^{(1)}
      \frac{\lambda^2}{E_\pi}  H_{22}^{(2)} \, \eta \, H_{20}^{(2)}
\; + \;  \frac{1}{2} (1 + 2 \alpha_5 ) \, H_{21}^{(1)} \frac{\lambda^1}{E_\pi^2}
      H_{22}^{(2)} \frac{\lambda^1}{E_\pi}  H_{21}^{(1)} \, \eta \,  H_{20}^{(2)}  \nn
&& {} - H_{21}^{(1)} \frac{\lambda^1}{E_\pi^2}
      H_{20}^{(2)} \, \lambda^1 \,   H_{21}^{(1)} \frac{\lambda^2}{E_\pi}  H_{22}^{(2)} 
\; - \;  2 \alpha_5 \,  H_{22}^{(2)} \frac{\lambda^2}{E_\pi^2}  H_{21}^{(1)} 
 \frac{\lambda^1}{E_\pi}  H_{21}^{(1)} \, \eta \, H_{20}^{(2)}  \nn
&& {} +  \frac{1}{2} (1 - 2 \alpha_5 ) \, H_{22}^{(2)} \frac{\lambda^2}{E_\pi}
      H_{21}^{(1)} \frac{\lambda^1}{E_\pi^2}  H_{21}^{(1)} \, \eta \,  H_{20}^{(2)}
\; - \;  \frac{1}{2} H_{21}^{(1)} \frac{\lambda^1}{E_\pi^3}
      H_{21}^{(1)} \, \eta \,   H_{20}^{(2)}\, \eta \,  H_{20}^{(2)} \nn
&& {} +  H_{21}^{(1)} \frac{\lambda^1}{E_\pi^3} H_{20}^{(2)} \, \lambda^1  H_{21}^{(1)}\,
\eta \,  H_{20}^{(2)}
\; - \;  \frac{1}{2} H_{21}^{(1)} \frac{\lambda^1}{E_\pi^3}
      H_{20}^{(2)} \, \lambda^1  \, H_{20}^{(2)}\, \lambda^1 \,  H_{21}^{(1)} 
 \bigg]  \eta +  \mbox{h.c.} \,,
\eeqa
\item{retardation corrections $\propto g_A^2 C_{S,T}/m$:}
\beqa
V &=& \eta \bigg[ \frac{1}{2}  \, H_{21}^{(1)} \frac{\lambda^1}{E_\pi} 
H_{40}^{(2)} \frac{\lambda^1}{E_\pi^2}  H_{21}^{(1)}  \,  \eta \, H_{20}^{(2)}  
\; - \;   H_{21}^{(1)} \frac{\lambda^1}{E_\pi} H_{40}^{(2)} \frac{\lambda^1}{E_\pi^2}
H_{20}^{(2)}  \, \lambda^1 \,  H_{21}^{(1)}  
\;   + \; \frac{1}{2}  \, H_{21}^{(1)} \frac{\lambda^1}{E_\pi^2} 
H_{40}^{(2)} \frac{\lambda^1}{E_\pi}  H_{21}^{(1)}  \, \eta \,  H_{20}^{(2)}   \nn [2pt]
&& {} - \left( \frac{1}{2} - \bar \beta_8 \right)  \,   H_{21}^{(1)} \frac{\lambda^1}{E_\pi^3} H_{21}^{(1)}  \, \eta \,
H_{20}^{(2)}  \, \eta \,  H_{40}^{(2)} 
\; + \;   H_{21}^{(1)} \frac{\lambda^1}{E_\pi^3} H_{20}^{(2)}  \, \lambda^1 \, H_{21}^{(1)} \eta \,
H_{40}^{(2)} \nn [2pt]
&& {}  -  \left( \frac{1}{2} + \bar \beta_8 \right)  \, H_{40}^{(2)} \, \eta \,  H_{21}^{(1)}
\frac{\lambda^1}{E_\pi^3} H_{21}^{(1)}  \, \eta \,  
H_{20}^{(2)}   \bigg]  \eta +  \mbox{h.c.} \,, 
\eeqa
\item{terms involving $1/m$-corrections to the $g_A$-vertex,  $\propto g_A^4/m$:}
\beqa
V &=& \eta \bigg[  \left( \frac{1}{2} + \bar \beta_9 \right) \, H_{21}^{(1)} \frac{\lambda^1}{E_\pi} H_{21}^{(1)} \, \eta
\, H_{21}^{(1)} \frac{\lambda^1}{E_\pi^2} H_{21}^{(3)} 
\; + \;  \left( \frac{1}{2} - \bar \beta_9 \right) \, H_{21}^{(1)} \frac{\lambda^1}{E_\pi} H_{21}^{(1)} \, \eta
\, H_{21}^{(3)} \frac{\lambda^1}{E_\pi^2} H_{21}^{(1)} \nn[2pt]
&&{} -  H_{21}^{(1)} \frac{\lambda^1}{E_\pi} H_{21}^{(1)}  \frac{\lambda^2}{E_\pi} 
 H_{21}^{(1)} \frac{\lambda^1}{E_\pi} H_{21}^{(3)}  
\;   -\;  H_{21}^{(1)} \frac{\lambda^1}{E_\pi} H_{21}^{(1)}  \frac{\lambda^2}{E_\pi} 
 H_{21}^{(3)} \frac{\lambda^1}{E_\pi} H_{21}^{(1)} 
\; + \;  \frac{1}{2} \, H_{21}^{(1)} \frac{\lambda^1}{E_\pi^2} H_{21}^{(1)} \, \eta
\, H_{21}^{(1)} \frac{\lambda^1}{E_\pi} H_{21}^{(3)} \nn [2pt]
&& {} +   \frac{1}{2} \, H_{21}^{(1)} \frac{\lambda^1}{E_\pi^2} H_{21}^{(1)} \, \eta
\, H_{21}^{(3)} \frac{\lambda^1}{E_\pi} H_{21}^{(1)} \bigg]  \eta +  \mbox{h.c.} \,,
\eeqa
\item{terms involving $1/m$-corrections to the $g_A$-vertex,  $\propto
    g_A^2 C_{S,T}/m$:}
\beq
V = \eta \bigg[  H_{21}^{(1)} \frac{\lambda^1}{E_\pi} H_{40}^{(2)}
\frac{\lambda^1}{E_\pi} H_{21}^{(3)} 
\; - \;  \left( \frac{1}{2}  - \bar \beta_9 \right) \, H_{21}^{(1)} \frac{\lambda^1}{E_\pi^2} H_{21}^{(3)} \, \eta \,
H_{40}^{(2)} 
\; - \;  \left( \frac{1}{2} + \bar \beta_9 \right) \, H_{21}^{(3)} \frac{\lambda^1}{E_\pi^2} H_{21}^{(1)} \, \eta \,
H_{40}^{(2)} \bigg]  \eta +  \mbox{h.c.} \,.
\eeq
\end{itemize}
Here and in what follows, we adopt the notation of
Refs.~\cite{Epelbaum:2007us,Bernard:2007sp, Kolling:2009iq,Kolling:2011mt}. In particular, the subscripts $a$ and $b$ in
$H_{ab}^{(\kappa)}$  refer to the
number of the nucleon and pion fields, respectively, while the
superscript $\kappa$ gives the inverse mass dimension of the
corresponding coupling constant\footnote{For $1/m$-corrections,
  $\kappa_i$ corresponds to the inverse power of coupling constants
  plus twice the power of $m^{-1}$.  In particular, $\kappa=2$ for the
nucleon kinetic energy term $H_{20}$.},
 see Ref.~\cite{Epelbaum:2007us} for more details. 
The chiral
order associated with a given contribution can easily be read off by adding together the dimensions
$\kappa$  of $H_{ab}^{(\kappa)}$. More precisely, it is given by $\sum_i \kappa_i -2$.  
In the above equations,  $\eta$ ($\lambda$)  denote 
projection operators onto the purely nucleonic (the remaining) part of the
Fock space satisfying $\eta^2 = \eta$, $\lambda^2 = \lambda$, $ \eta \lambda 
= \lambda \eta = 0$ and $\lambda + \eta = {\bf 1}$. The superscript
$i$ of $\lambda^i$ refers to  the number of pions in the
corresponding intermediate state. Further, $E_\pi$ denotes the total
energy of the 
pions in the corresponding state, $E_\pi = \sum_i \sqrt{\vec l_i\, ^2 +
  M_\pi^2}$, with $\vec l_i$ the corresponding pion momenta.

\def\theequation{\Alph{section}.\arabic{equation}}
\setcounter{equation}{0}
\section{Coordinate-space representation}
\label{app2}

In this appendix we give the coordinate-space representation of the
obtained 3NF contributions. Here and in what follows, we adopt the
notation of Refs.~\cite{Epelbaum:2007us,Bernard:2007sp} and use the dimension-less 
profile functions $U_{1,2}$, $W$ and $W_{1,2}$ defined for a general form of a
local regulator  function $F^\Lambda$  according to 
\beqa
U_1 (x) &=& \frac{4 \pi }{M_\pi} \int \frac{d^3 q}{(2 \pi )^3} \, \frac{e^{i
    \vec q \cdot \vec x/M_\pi}}{\vec q \, ^2 + M_\pi^2} \, F^\Lambda (q)
\stackrel{\Lambda \to \infty}{\longrightarrow} \frac{e^{-x}}{x}\,, \nn
U_2 (x) &=& 8 \pi M_\pi \, \int \frac{d^3 q}{(2 \pi )^3} \, \frac{e^{i
    \vec q \cdot \vec x/M_\pi}}{(\vec q \, ^2 + M_\pi^2)^2} \, F^\Lambda (q)
\stackrel{\Lambda \to \infty}{\longrightarrow} e^{-x}\,, \nn
W (x) &=& \frac{1}{M_\pi^3} \,
  \int \frac{d^3 q}{(2 \pi )^3} \, e^{i
    \vec q \cdot \vec x/M_\pi} \, F^\Lambda (q)
\stackrel{\Lambda \to \infty}{\longrightarrow} \delta^3 (x) \,,
\nn
W_1(x) &=& \frac{4 \pi }{M_\pi^2} \int \frac{d^3 q}{(2 \pi )^3} \, e^{i
    \vec q \cdot \vec x/M_\pi}\, A (q)  \, F^\Lambda (q)
  \stackrel{\Lambda \to \infty}{\longrightarrow}   \frac{e^{-2 x}}{2 x^2}\,, \nn
W_2(x) &=& \frac{4 \pi }{M_\pi^4}  \int \frac{d^3 q}{(2 \pi )^3} \, e^{i
    \vec q \cdot \vec x/M_\pi}\, q^2 A (q)  \, F^\Lambda (q) = - \nabla_x^2  W_1 (x)
\,. 
\eeqa

For the two-pion-exchange-contact terms in
Eqs.~(\ref{2picont1},\ref{2picont2}) we find:
\beqa
V_{\rm 2\pi-cont} &=&  \frac{g_A^2 C_T M_\pi^7}{ 192 \pi^2 F_\pi^4} 
\Big\{ 2 \fet \tau_1 \cdot \fet \tau_2 (\vec \sigma_2 \cdot \vec
\sigma_3 ) \Big[ 4 \pi (3 g_A^2 - 1) W(x_{12}) - 2 g_A^2 U_1 (2
x_{12}) + (2 g_A^2 - 1) (2 W_1 (x_{12}) + W_2 (x_{12})) \Big] \nn
&& {} - 9 g_A^2 \Big[ (\vec \sigma_1 \cdot \vec \nabla_{12}) (\vec \sigma_2
\cdot \vec \nabla_{12}) W_1 (x_{12}) + (\vec \sigma_1 \cdot \vec
\sigma_2 ) W_2 (x_{12}) \Big] \Big\} W (x_{32}) \; +\;\mbox{5
  permutations} \,.
\eeqa

Next, the coordinate-space representation of the 
relativistic corrections to the two-pion exchange 3NF emerges
from taking the Fourier transform of Eq.~(\ref{2pi_rel_mom}):
\beq 
\label{2pi_rel_coord}
V_{2\pi , \; 1/m} = \frac{g_A^2 M_\pi^6}{1024 m \pi^2 F_\pi^4} 
\, \Big( \fet \tau_1 \cdot \fet \tau_3 \, \tilde A_{123} + [ \fet \tau_1 \times
\fet \tau_2] \cdot \fet \tau_3 \, \tilde B_{123} \Big)\; +\;\mbox{5
  permutations} \,, 
\eeq
where 
\beqa
\tilde A_{123} &=& ( \vec \sigma_1 \cdot \vec \nabla_{12}  ) ( \vec \sigma_3 \cdot \vec
\nabla_{32}  ) \, \tilde a_{123} + \Big\{ \Big( \vec \sigma_3 \cdot 
 \frac{\vec p_3 }{M_\pi}   \Big), \;  ( \vec \sigma_1 \cdot \vec
\nabla_{12}  )  \, \tilde c_{123} \Big\} \,, \nn
\tilde B_{123} &=& ( \vec \sigma_1 \cdot \vec \nabla_{12}  ) ( \vec \sigma_3 \cdot \vec
\nabla_{32}  ) \, \tilde  b_{123} + \Big\{ \Big( \vec \sigma_3 \cdot \frac{\vec p_3}{M_\pi}  \Big),
\;  ( \vec \sigma_1 \cdot \vec
\nabla_{12} )  \, \tilde d_{123} \Big\}\,,
\eeqa
and the functions $\tilde a_{123}$,   $\tilde b_{123}$,   $\tilde
c_{123}$ and   $\tilde d_{123}$
are given by:  
\beqa
\label{rspace}
\tilde a_{123}&=& g_A^2 (1 - 2 \bar \beta_8 ) (\vec \nabla_{12} \cdot
\vec \nabla_{32})^2 \, U_2 (x_{12}) \, U_1 (x_{32}) \nn 
&+& g_A^2 \Big\{ (1 - 2 \bar \beta_8 ) \frac{\vec p_2}{M_\pi} + (1 + 2 \bar \beta_8 )
\frac{\vec p_1}{M_\pi} , \; \vec \nabla_{12} \,  [ \vec \nabla_{12} \times \vec
\nabla_{32}] \cdot \vec \sigma_2 \, U_2 (x_{12}) \, U_1 (x_{32})
\Big\} - 2 g_A^2 (2 \bar \beta_9 - 1) \nabla_{12}^2 \, U_1 (x_{12}) \, 
U_1 (x_{32} ) \nn
&+& 2 g_A^2 (2 \bar \beta_9 - 1) \Big\{ \frac{\vec p_2}{M_\pi} , \;
[\vec \sigma_2 \times \vec \nabla_{12} ] \, U_1 (x_{12}) \, U_1
(x_{32}) \Big\}\,, \nn
\tilde c_{123} &=& - 2 g_A^2 (2 \bar \beta_9 + 1) [ \vec \nabla_{12}
\times \vec \nabla_{32}] \cdot \vec \sigma_2 \, U_1 (x_{12}) \, U_1
(x_{32}) \,, \nn 
\tilde b_{123} &=&  g_A^2 (1 - 2 \bar \beta_8 ) [\vec \nabla_{12} \times
\vec \nabla_{32} ] \cdot \vec \sigma_2 (\vec \nabla_{12} \cdot \vec
\nabla_{32}) \, U_2 (x_{12}) \, U_1 (x_{32}) \nn
&-& g_A^2 \Big\{ (1 - 2 \bar \beta_8 ) \frac{\vec p_2}{M_\pi} + (1 + 2 \bar \beta_8 )
\frac{\vec p_1}{M_\pi} , \; \vec \nabla_{12} \,  ( \vec \nabla_{12} \cdot \vec
\nabla_{32} ) \, U_2 (x_{12}) \, U_1 (x_{32}) \Big\}  - 2 \Big\{ \frac{\vec p_3}{M_\pi} -
\frac{\vec p_2}{M_\pi} , \; \vec \nabla_{32} \, U_1 (x_{12}) \, U_1 (x_{32}) 
\Big\} \nn
&-& 2 g_A^2 (2 \bar \beta_9 - 1) \Big\{ \frac{\vec p_2}{M_\pi} , \; \vec \nabla_{12}
\,  U_1 (x_{12}) \, U_1 (x_{32}) \Big\}  - 2 [\vec \nabla_{12} \times \vec
\nabla_{32}]
\cdot \vec \sigma_2 \, U_1 (x_{12}) \, U_1 (x_{32})\,, \nn
\tilde d_{123} &=&  2 g_A^2 (2 \bar \beta_9 + 1) ( \vec \nabla_{12}
\cdot \vec \nabla_{32})  \, U_1 (x_{12}) \, U_1
(x_{32}) \,, 
\eeqa
where $\vec x_{ij} \equiv M_\pi \vec r_{ij}$, $x_{ij} \equiv | \vec
x_{ij} |$ and $\vec r_{ij} = \vec r_i - \vec r_j$ is the distance
between the nucleons $i$  and $j$. Further, the $\vec \nabla_{ij}$ act
on $\vec x_{ij }$ (i.e.~are dimension-less)  and $\vec p_i$ refer 
to the momentum operator of the nucleon $i$.  Curly brackets denote,
as usual, the anti-commutator of two operators, $\{X, \; Y \} \equiv
XY + YX$, and   $\{ \vec X, \; \vec Y \} \equiv
\vec X \cdot \vec Y + \vec Y \cdot \vec X$.  

Finally, the $1/m$-corrections to the  one-pion-exchange-contact
topology have the form 
\beq
 \label{1pi_cont_rel_coord}
V_{1\pi{\rm -cont}, \; 1/m} = \frac{g_A^2 M_\pi^6}{64 m \pi F_\pi^2} \, 
\fet \tau_1 \cdot \fet \tau_2  \Big( (\vec \sigma_1 \cdot \vec
\nabla_{12}) \, 
\tilde f_{123} + \Big\{ \vec \sigma_1 \cdot \frac{\vec p_1}{M_\pi}, \; \tilde
g_{123} \Big\} \Big)\; +\;\mbox{5
  permutations} \,,
\eeq
with the functions    $\tilde
f_{123}$ and   $\tilde g_{123}$
given by:  
\beqa
\tilde f_{123} &=& (1 - 2 \bar \beta_8 ) (\vec \nabla_{12} \cdot \vec
\nabla_{32})
\Big( C_S (\vec \nabla_{12} \cdot \vec \sigma_2 ) + C_T (\vec
\nabla_{12} \cdot \vec \sigma_3 ) \Big) U_2 (x_{12}) \, W (x_{32}) \nn
&-& C_T \Big\{ (1 - 2 \bar \beta_8 ) \frac{\vec p_2}{M_\pi} + (1 + 2 \bar \beta_8 )
\frac{\vec p_1}{M_\pi} , \; \vec \nabla_{12} \, \vec \nabla_{12} \cdot [\vec
\sigma_2 \times \vec \sigma_3 ] \, U_2 (x_{12}) \, W (x_{32}) \Big\} \nn
&-& 2 (2 \bar \beta_9 - 1) \Big( C_S (\vec \nabla_{32} \cdot \vec \sigma_2 ) + C_T (\vec
\nabla_{32} \cdot \vec \sigma_3 ) \Big) U_1 (x_{12}) \, W (x_{32}) 
+ 2 C_T (2 \bar \beta_9 - 1) \Big\{ \frac{\vec p_2}{M_\pi} , \; 
[\vec \sigma_2 \times \vec \sigma_3 ] U_1 (x_{12}) W(x_{32}) \Big\}\,,
\nn
\tilde g_{123} &=& - 2 C_T (2 \bar \beta_9 + 1 ) \vec \nabla_{12}
\cdot 
[\vec \sigma_2 \times \vec \sigma_3 ] \, U_1 (x_{12}) \, W (x_{32})\,.
\eeqa


\end{document}